# Effects of alpha stopping power modelling on the ignition threshold in a directly-driven Inertial Confinement Fusion capsule

M. Temporal[1, a], B. Canaud[2], W. Cayzac[2], R. Ramis[3], and R.L. Singleton Jr[4, b]

*1) Centre de Mathématiques et de Leurs Applications, ENS Cachan and CNRS, 61 Av. du President Wilson, F-94235 Cachan Cedex, France*

*2) CEA, DIF, F-91297 Arpajon Cedex, France*

*3) ETSI Aeronáutica y del Espacio, Universidad Politécnica de Madrid, 28040 Madrid, Spain*

*4) Los Alamos National Laboratory, Los Alamos, New Mexico 87545, USA*

**Abstract.** The alpha-particle energy deposition mechanism modifies the ignition conditions of the thermonuclear Deuterium-Tritium fusion reactions, and constitutes a key issue in achieving high gain in Inertial Confinement Fusion implosions. One-dimensional hydrodynamic calculations have been performed with the code Multi-IFE [R. Ramis and J. Meyer-ter-Vehn, Comp. Phys. Comm. **203**, 226 (2016)] to simulate the implosion of a capsule directly irradiated by a laser beam. The diffusion approximation for the alpha energy deposition has been used to optimize three laser profiles corresponding to different implosion velocities. A Monte-Carlo package has been included in Multi-IFE to calculate the alpha energy transport, and in this case the energy deposition uses both the LP [C.K. Li and R.D. Petrasso, Phys. Rev. Lett. **70**, 3059 (1993)] and the BPS [L.S. Brown, D.L. Preston, and R.L. Singleton Jr., Phys. Rep. **410**, 237 (2005)] stopping power models. Homothetic transformations that maintain a constant implosion velocity have been used to map out the transition region between marginally-igniting and high-gain configurations. The results provided by the two models have been compared and it is found that - close to the ignition threshold - in order to produce the same fusion energy, the calculations performed with the BPS model require about 10% more invested energy with respect to the LP model.

[a] e-mail: mauro.temporal@hotmail.com

[b] LA-UR-17-20568

## 1 Introduction

In Inertial Confinement Fusion (ICF) [1, 2], the Deuterium-Tritium (DT) thermonuclear fuel is compressed and heated in order to ignite the fusion reaction D + T → α + n + Q, where Q = 17.6 MeV. In the central ignition scheme, the implosion of a spherical capsule aims to compress and heat ($T_h \approx 10$ keV) a small portion of the fuel called the hot-spot, a region where the fusion reactions take place. In successful implosions, the relatively large fuel areal density ($\rho r \geq 3$ g/cm$^2$) allows the propagation of a thermonuclear burn wave through a significant portion of fuel, providing high fractional burn-up ($\approx 1/3$) and high energy gain G ≈ 100.

The kinetic energy of the α-particles produced by the DT fusion reaction is $\varepsilon_0 = Q/5$ (by momentum conservation, 4Q/5 goes to the neutron kinetic energy). Here, it is assumed that the neutrons escape from the fuel while the α-particles self-heat the hot-spot mass, and then progressively propagates the fusion reactions throughout the surrounding fuel. To optimize this process, the range of the alpha particles ($R_\alpha$) should match the hot-spot dimension $R_\alpha \approx \rho_h r_h \approx 0.3$ g/cm$^2$. Thus, the energy needed to heat the hot-spot mass $(4\pi/3) r_h^3 \rho_h$ to the temperature $T_h$ scales as $T_h (r_h \rho_h) r_h^2$, which in turn is proportional to the alpha range. Of course, the longer the range of the α particle, the larger the heated mass and the required energy for the DT ignition. Therefore, the process of α-energy transport is crucial in the determination of the DT ignition conditions.

To address this topic, we used the one-dimensional (1D) hydrodynamic code Multi-IFE [3] to simulate the implosion of an ICF capsule directly irradiated by a laser beam. In the baseline version of the code, the diffusion approximation is used to model the α-particle energy deposition, and for this work a 3D Monte-Carlo (MC) package has been developed to calculate the α-particle transport. The MC package also allows one to use different ion stopping power models, currently the LP (Li - Petrasso) [4] and BPS (Brown - Preston - Singleton) [5] models have been implemented in the MC package.

The basic idea of this work is to analyse how a more recent and updated model (BPS) can modify the results previously obtained from the well-established LP model. In the BPS model, a quasi-exact evaluation of the stopping power is achieved for weakly coupled plasmas over a wide range of coupling regimes (from moderate down to the weakly coupled regime). The model includes an exact treatment of the quantum-to-classical scattering transition. A preliminary study [6] showed that the α-particle ranges are larger with the BPS model (in comparison with the LP model), and this leads to higher invested energies. Here, we investigate the effect of this new model on kinetic ignition thresholds deduced from hydrodynamic scaling (homothetic transformations) [7].

The paper is organized as follows. First a suite of several thousand hydrodynamic calculations has been performed within the diffusive approximation with the laser pulse parameters randomly chosen. This allows to select the optimal laser pulses that maximize the thermonuclear energy fusion (Section 2). Section 3 gives some details of the Monte-Carlo package, and Sec. 4 is devoted to comparing the ignition curves obtained via homothetic transformations using the two stopping power models (LP, BPS), as well as a comparison with the diffusive approximation for the alpha energy deposition. Conclusions are draw in the last Section 5.

## 2 Capsule and laser pulse

The code Multi-IFE [3] has been used to simulate the hydrodynamic evolution of a capsule directly irradiated by a laser beam (3ω, $\lambda = 0.35$ μm). For simplicity, in these calculations the photons are considered to propagate in the radial directions (1D ray-tracing), thus the incident laser power equals the absorbed power (P), and the light absorption is modelled by inverse-bremsstrahlung. The code assumes two plasma temperatures, ionic and electronic, heat conduction with the flux harmonically limited to 8 %, multi-group radiation and tabulated QEOS [8, 9] equations of state. In the standard version of the Multi-IFE code alpha energy deposition is calculated in the diffusive approximation [10, 11].

A capsule [12, 13] designed for Direct Drive ICF has been considered. The capsule (see Fig. 1) has an outer radius of 815 μm, and it is composed of an external plastic ablator (CH, $\rho = 1.07$ g/cm$^3$) with a thickness of 24 μm that encloses the shell of solid cryogenic DT ($\rho = 0.25$ g/cm$^3$) of thickness $\Delta_{DT} = 198$ μm and mass $m_{DT} = 300$ μg. Hereafter we refer to the solid cryogenic DT as the payload mass. The inner sphere of radius r = 593 μm is filled by DT gas at density $\rho = 0.3$ mg/cm$^3$ and the initial aspect ratio is A = r / $\Delta_{DT}$ = 3.

In Fig. 1 is also shown the flow chart for a marginally igniting case corresponding to an implosion velocity V = 300 μm/ns. This implosion velocity has been defined as the maximum mean velocity - mass average - evaluated over the cells with negative speed (imploding cells). For this case the maximum kinetic energy ($E_K$) reached by the payload during the implosion is $E_K = 13.4$ kJ, while the total absorbed energy is $E_A = 330$ kJ and the output fusion energy is $E_F = 2.8$ MJ, thus the energy gain is G = $E_F$ /$E_A$ = 8.

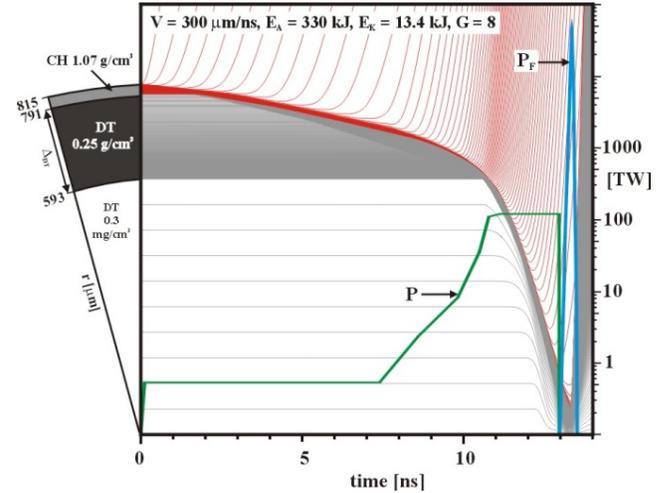

**Fig. 1.** Capsule design and temporal evolution of the Lagrangian cell interfaces. The green curve (P) is the absorbed laser power while the blue one ($P_F$) is the thermonuclear output fusion power.

In order to optimize the laser pulse, a set of calculations has been performed assuming randomly selected laser absorbed power. The laser pulse profile $P_i(t_i)$ is made by 8 points (i = 1, 8) in the t-P plane. The foot of the pulse is set to the constant power $P_1 = 0.6$ TW, while the main drive is flat ($P_7 = P_6$). An initial and final time-rise of 0.1 ns are used. The pulse is build with random powers $P_i$ and random times $t_i$ that lie inside selected zones defined by the shaded areas shown in Fig. 2.

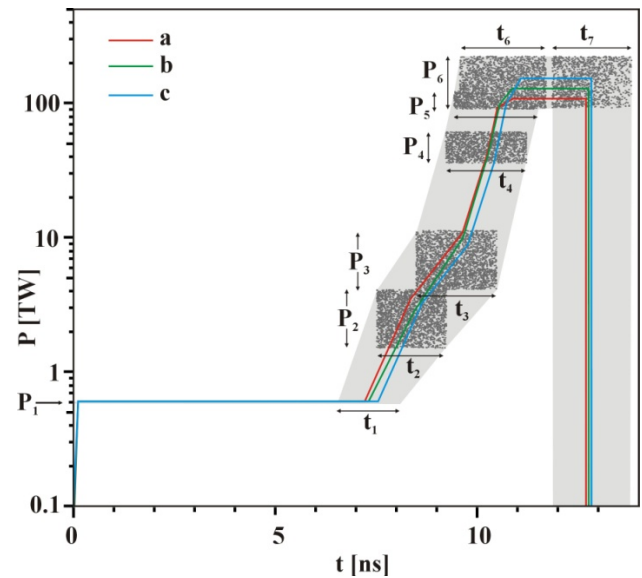

**Fig. 2.** Laser absorbed power (P) as a function of the time. The shaded areas delimit the parametric space of the randomly selected laser pulse $P_i(t_i)$.

For each numerical simulation, the code calculates the energy gain G and the implosion velocity V. The results are shown in the Fig. 3 where each dot represents a calculation in the parametric space [G, V]. It clearly delineates a non-igniting region (G < 1) for relatively small implosion velocity, and an ignition threshold that corresponds to an implosion velocity of about 300 μm/ns. Three cases that maximize the energy gain at given implosion velocities have been selected: a non-igniting case (a) corresponding to V = 280 μm/ns, a case (b) at the ignition threshold with V = 300 μm/ns, and a robust igniting case (c) with V = 320 μm/ns. The positions in the space [G, V] of these selected cases are shown in Fig. 3, and the corresponding laser pulses are shown by the blue (a), green (b) and red (c) curves in the Fig. 2. These three cases have been used in the section 4 to perform the homothetic study.

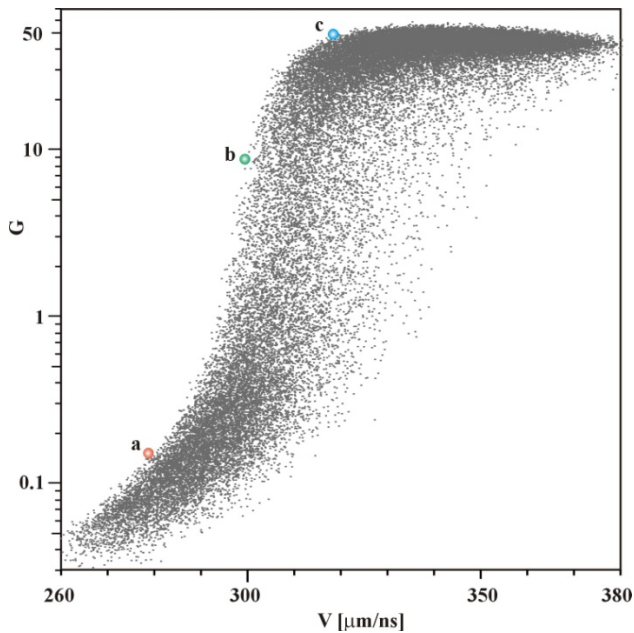

**Fig. 3.** Energy gain (G) as a function of the implosion velocity (V). Three cases (a, b and c) have been selected and correspond to the maximum gain at the implosion velocity of 280, 300 and 320 μm/ns.

## 3 Energy transport of the α-particles

The power density provided by the DT fusion reactions associated with the α-particles is given by $\varepsilon_0 \langle\sigma v\rangle n^2$ [3], where the reactivity $\langle\sigma v\rangle$ is calculated according to by Bosch and Hale [14] and n is the number density of the Deuterium and Tritium ions ($n = n_D = n_T$) for an equimolar DT plasma.

A diffusive model can be used to solve the α-particles energy deposition through the plasma. In this case a diffusive coefficient K, that multiply the gradient of the energy density of the α-particles, must be used. The diffusive coefficient K is proportional to the square of the initial α-particle velocity ($v_0 = 1.3 \times 10^9$ cm/s) and to the alpha-to-electron relaxation time ($\tau^{\alpha e}$): $K \propto \tau^{\alpha e} v_0^2$. The time $\tau^{\alpha e} \propto T_e^{3/2} / (\log\Lambda\, n_e)$ [2] - with $T_e$ being the electron temperature, $\log\Lambda$ the Coulomb logarithm and $n_e$ the electron number density - is a characteristic time of the energy deposition to the plasma and modulates the diffusive process. The solution of the diffusive equation only provides the specific power deposition to the plasma. This power heats both ion and electron populations, therefore another parameter is needed to separate the two contributions. The code uses the factor $F_e = 33\,\text{keV} / (33\,\text{keV} + T_e)$, introduced by Fraley *et al.* [15], to estimate the fraction of the α-particle energy deposited into electrons ($F_e$) and into ions (1 - $F_e$).

Another approach consists in estimating the α-to-plasma energy deposition by using a 3D Monte-Carlo (MC) package. In each hydro-cell and at every hydrodynamic time $t^n$, a given number $n_\alpha$ of α-particles proportionally to the produced thermonuclear power, are injected. Then, an initial random direction **u** is assigned to each α-particle, and it is assumed that they propagate along straight trajectories in the 3D space. By means of a stopping power model, the MC evaluates the energy deposition of the α-particles on the electron and ion populations: $(dE/dz)_e$ and $(dE/dz)_i$. The integral of the stopping power along the particle path provides the energy deposited in each cell. These terms are evaluated in each hydrodynamic time-step; then, the α-particle's position, direction and energy are stored and the process restarts in the next time-step using the new plasma parameters. In this way, at each time-step, the code follows the newly generated α-particles as well as the previous ones that have not yet been thermalized in the plasma. The α-particle is no-longer followed when its energy $\varepsilon$ becomes smaller than $\varepsilon_0/1000$ or $\varepsilon < \text{Min}\,[\,T_e, T_i\,]$, and the residual energy is deposited locally. The stopping powers of the electrons $(dE/dz)_e$ and of the ions $(dE/dz)_i$ have been calculated by means of two models: the Li - Petrasso (LP) [4] and the Brown - Preston - Singleton (BPS) [5].

In a typical calculation, the number of α-particle paths, over which the MC samples the energy deposition, exceed $10^5$ (as order of magnitude: 100 cells x 100 time-steps x 10 α-particles), and the stopping power calculations are a factor hundred bigger. As a consequence, the CPU time devoted to solve the BPS stopping power model becomes too large. To overcome this drawback, tabulated values of the electronic $(dE/dz)_e$ and ionic $(dE/dz)_i$ stopping powers have been employed. For both models (LP and BPS), stopping power tables with $(dE/dz)_e$ and $(dE/dz)_i$ as a function of the DT plasma density ($\rho$), temperature ($T = T_e = T_i$), and α-particle energy ($\varepsilon$), have been generated. The tables are built with 72 values of densities (from $10^{-4}$ g/cm$^3$ to $10^4$ g/cm$^3$, with 9 points per decade), 54 values of temperatures (from 1 eV to 1 MeV), and 100 values for the particle energy that follow a geometrical progression from 1.5 keV up to the initial energy $\varepsilon_0 = 3.52$ MeV. It has been verified that the stopping power evaluated by solving directly the models and the solution provided by a linear interpolation of the tabulated values do not differ more than 1 %.

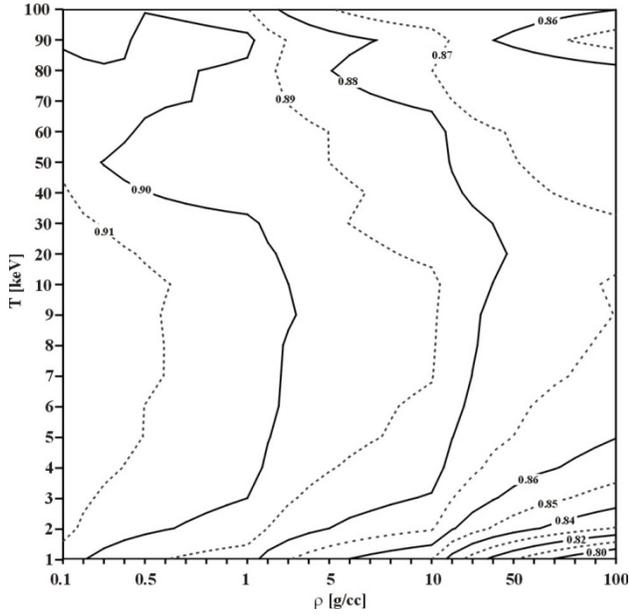

**Fig. 4.** Ratio ($R^{LP} / R^{BPS}$) of the α-particle range evaluated with the LP and BPS models as a function of density (ρ) and temperature (T = $T_e$ = $T_i$) of a DT plasma.

For an α-particle with initial energy $\varepsilon_0$, the range ($R^{BPS}$) evaluated with the BPS model is larger than the one ($R^{LP}$) calculated using the LP model. Fig. 4 shows the ratio $R^{LP} / R^{BPS}$ as a function of the density and temperature for an equimolar DT plasma. It is found that the ranges $R^{BPS}$ are always 10 % larger than the corresponding ranges $R^{LP}$. It is worth noticing that the accurate solution of the BPS model provides even larger range ratios [6]. Nevertheless, in typical ICF hot-spot conditions, this additional lengthening of the range only concerns the energy deposition of the last keV of the α-particle energy just above the plasma temperature. Thus, close to the α-particle thermalization, the error in the energy deposition introduced by using the tabulated stopping powers amounts to about 1/3520 of the total and does not modify significantly the overall ignition process.

The increase in the α-particle range predicted by the BPS model has a sizable effect on the ignition of the DT nuclear fusion reactions and leads to more severe ignition conditions, as it is shown in the next section 4.

## 4 Homothetic analysis

The homothetic transformation technique enables to carry out extensive studies of the effects of the physical parameters over a broad variety of configurations without the need of an expensive optimization. Homothetic transformations, known also as hydrodynamic scaling, are used to scale the parameters of a given reference capsule and laser pulse in order to generate a family of cases with similar hydrodynamic evolution. The transformation assumes that capsule dimensions and time-lengths of the laser pulse scale linearly with a homothetic factor $f$ while the laser power scales as $f^2$, consequently mass and energies scales as $f^3$ while velocities are kept constant.

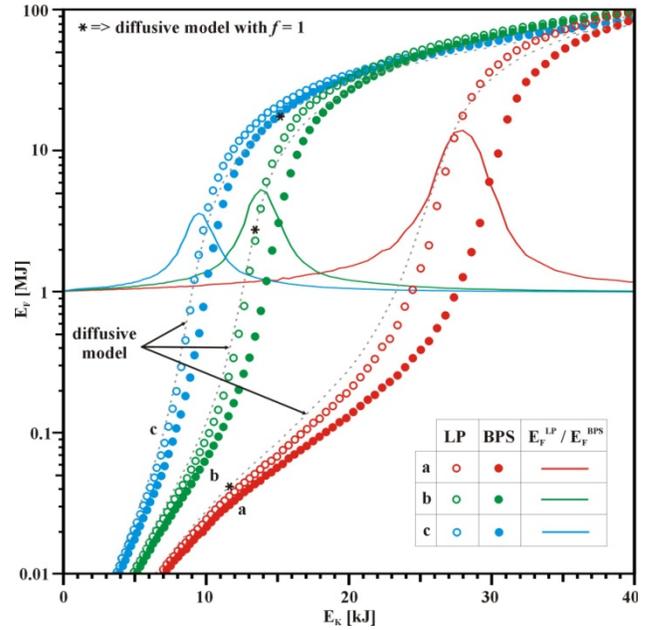

**Fig. 5.** Fusion energy $E_F$ as a function of the payload kinetic energy $E_K$. Full (void) circles refer to the calculations with the BPS (LP) stopping power model, whilst dashed curves use the diffusive approximation. The solid lines show the ratio between the fusion energies $E_F^{LP} / E_F^{BPS}$ provided by using the two models.

We considered the capsule design defined in Fig. 1 and the three laser-pulses (a, b, and c) shows in Fig. 2, corresponding to implosion velocities $V_a = 280$ μm/ns, $V_b = 300$ μm/ns, and $V_c = 320$ μm/ns. Homothetic transformations ($r \propto f$, $t \propto f$, and $P \propto f^2$) have been applied to these three cases. The variation of the scale factor $f$ allows us to compare cases that have the same implosion velocity with the reference cases a, b, and c for which $f = 1$. Two sets of calculations have been performed. The first set uses the Monte-Carlo α-particle transport associated with the LP stopping power model, and the second set uses the BPS model. Numerical calculations using the diffusive approximation model for the energy deposition of α-particles have also been performed. The thermonuclear output fusion energy $E_F$ and the maximum payload kinetic energy $E_K$ have been computed for each calculation. Such an approach allows one to scan the parametric space between marginally to fully igniting and burning designs. A kinetic energy threshold can then be defined as the transition between the two asymptotic regimes (non burning and burning cases). This threshold is very sensitive to differences between the models used in the calculations.

The results have been summarized in Fig. 5. Results obtained using the diffusive approximation (dashed curves) approximately match with the LP data (void circles). Indeed, the diffusive coefficient and the Fraley factor have been tuned in accord with classical stopping power models like LP. In contrast, for a fixed given output fusion energy $E_F$, the calculations using the BPS stopping power model require a larger payload kinetic energy in comparison to the cases using the LP model. The larger differences occur near to the ignition threshold where the calculations performed with the LP model predict significantly larger fusion energies ( $E_F^{LP} / E_F^{BPS} > 1$ ) as shown by the solid lines in Fig. 5. Moreover, the increment of the fusion energy at the ignition threshold is larger for smaller implosion velocities. Indeed, a maximum increase ($E_F^{LP} / E_F^{BPS}$) of a factor 14, 5.5 and 3.6 is found for the cases with implosion velocity $V_a = 280$ μm/ns, $V_b = 300$ μm/ns, and $V_c = 320$ μm/ns, respectively. The increase in the required maximum payload kinetic energy, $E_K$, corresponds to an analogous increase of the absorbed energy $E_A$ that, for this target, scale as $E_A \approx 25\, E_K$.

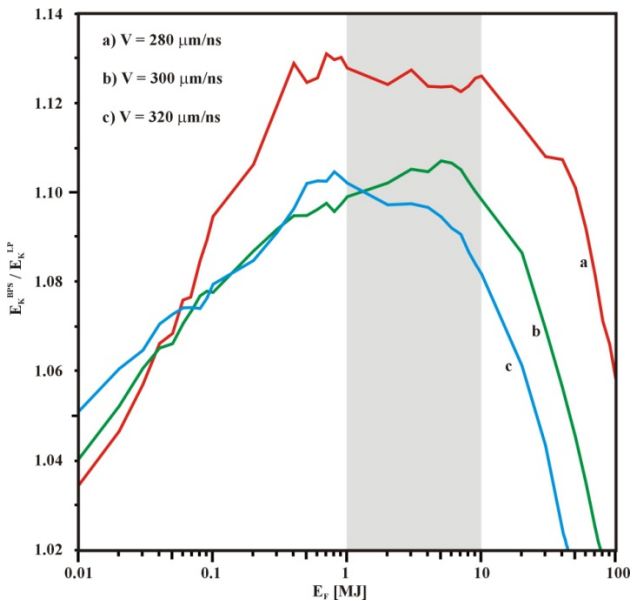

**Fig. 6.** Ratio of the payload kinetic energies $E_K^{BPS} / E_K^{LP}$ as a function of the output fusion energy, $E_F$.

The data of Fig. 5 have been used to evaluate the payload kinetic energy, $E_K$, required for obtaining a given fusion energy $E_F$. A linear fit of the fusion energies has been used to evaluate the functions $E_K^{LP}(E_F)$ and $E_K^{BPS}(E_F)$. The ratio $E_K^{BPS} / E_K^{LP}$ is shown in Fig. 6 as a function of the fusion energy $E_F$. It is found that around the ignition threshold - 1 MJ < $E_F$ < 10 MJ - the required payload kinetic energy is about 10 % larger using the BPS model instead of the LP model. These results are compatible with those shown in Ref. [6] where - for an idealized hot-spot characterized by a mass $m_h = 100$ μg and a uniform temperature $T_h = 3$ keV - an increase of the required hot-spot areal density $\rho_h r_h$ of approximately 10 % were estimated.

## 5 Conclusions

An ICF capsule with an initial aspect ratio A = 3 has been considered. The 1D hydro-radiative code Multi-IFE has been used to optimize the absorbed laser pulses for three reference cases characterized by implosion velocities of 280, 300 and 320 μm/ns. The hydrodynamic code uses a Monte-Carlo package that calculates the α-particle energy deposition by means of two different stopping power models (Li - Petrasso and Brown - Preston - Singleton). Homothetic transformations of the reference cases have been performed, providing the ignition curves in terms of the output fusion energy as a function of the maximum payload kinetic energy.

The results provided using the two models (LP and BPS) have been compared, and it has been found that the ignition curves shift to higher energies using the BPS model. Significant differences appear close to the ignition threshold, where the required invested energies increase by about 10% with the BPS model in comparison to the LP model. Finally, it is worth noting that this effect may adversely affect the robustness of existing designs and should be taken into account in the design of capsule configurations that work close to the ignition threshold.

M.T. and B.C. express their thank to Daniel Bouche for the support given to this work. R.R. has been supported by the Spanish MINECO project ENE2014-54960-R6065.